# Impact of hydrogen incorporation on electronic and magnetic structure of X2CrNi18-9 stainless steel

Torben Tappe[a,b], Louis Becker[c], Gaurav Kanu[d], Thomas F. Headen[e], Dirk Honecker[e], Gabi Schierning[a,b,f], Santiago Benito[c], Sebastian Weber[c], Klara Lünser[a,b], Sabrina Disch[d,f,]*

a) Applied Quantum Materials, Institute for Energy und Materials Processes (EMPI), Faculty of Engineering, University of Duisburg-Essen, 47057 Duisburg, Germany

b) Research Center Future Energy Materials and Systems (RC FEMS), University Alliance Ruhr, 44780 Bochum, Germany

c) Chair of Materials Technology, Ruhr University Bochum, Universitaetsstr. 150, 44801 Bochum, Germany

d) Faculty of Chemistry, University of Duisburg-Essen, Universitätsstr. 5-7, 45141 Essen, Germany

e) ISIS Neutron and Muon Source, Science and Technology Facilities Council, Rutherford Appleton Laboratory, Didcot OX11 0QX, United Kingdom

f) Center for Nanointegration Duisburg-Essen (CENIDE), University of Duisburg-Essen, Universitätsstr. 5-7, 45141 Essen, Germany

* Corresponding author: Sabrina Disch, sabrina.disch@uni-due.de

**Abstract**

Hydrogen absorption significantly alters the mechanical properties of steel. However, absorbed hydrogen also influences its electronic and magneto-structural properties, helping to interpret how hydrogen is incorporated. This study therefore investigates the influence of hydrogen incorporation on the electronic and magneto-structural properties of X2CrNi18-9 stainless steel in different microstructural states. Microstructural characterization included analytic electron microscopy mapping, X-ray diffraction and thermodynamic stability maps to evaluate grain size, dislocation density and chemical homogeneity. The electronic properties were characterized using the Seebeck coefficient, while the magneto-structural properties were investigated using diffuse neutron scattering and small-angle neutron scattering (SANS). Hydrogen incorporation showed clear changes in the Seebeck coefficients. Magnetic SANS in conjunction with diffuse neutron scattering indicates the existence of nanoscale inhomogeneities with the same fcc structure as the bulk, but with correlation lengths of a few nanometres. The size of these inhomogeneities increased with hydrogen incorporation, suggesting that hydrogen preferentially accumulates in their vicinity. However, no direct correlation between the electronic and magneto-structural properties and the dislocation density could be demonstrated. We suggest that studies such as these will lead in the medium term to the development of guidelines for material design to make steels more resistant to hydrogen.

1. Introduction

Hydrogen is one of the most important building blocks of the energy transition [1]. The industrial use of hydrogen as an energy carrier and reducing agent is steadily increasing [2]. Therefore, all materials that come into contact with hydrogen must be adapted for this application [3]. Hydrogen embrittlement of steels is a particularly major hurdle toward widespread adoption in this context, and materials with high resistance to hydrogen embrittlement are being developed for this purpose [4-9]. Several aspects are particularly important when selecting or developing steels for hydrogen applications [10]. Austenitic steel is known for its high hydrogen resistance over a wide range of operation conditions and is, for certain applications, preferred over ferritic or alpha-martensitic steels [11]. This is due to the higher solubility of hydrogen in austenite compared to Fe-base bcc structures, as well as its low hydrogen diffusivity [12]. Chemical inhomogeneities and defects influence hydrogen absorption, segregation, and trapping in the material [13]. Therefore, the processing of the material and its homogenization e.g., by thermal post-treatment are crucial for the sensitivity to hydrogen embrittlement [14, 15]. It has been shown recently that austenitic stainless steel produced by a laser powder bed fusion (PBF-LB/M) process may exhibit higher resistance to hydrogen embrittlement than conventionally processed material of the same composition [14, 16]. Besides grain size, chemical homogeneity was identified as one of the decisive factors which was better in the PBF material than in the conventionally processed reference [16]. The improved chemical homogeneity of additively manufactured austenitic stainless steels reduces their tendency to form martensite upon cryogenic cooling and/or deformation [16, 17]. This increase in local austenite stability is a decisive factor expected to contribute to the higher hydrogen embrittlement resistance of additively manufactured powder metallurgical parts [16, 18]. In addition, defect densities, particularly the dislocation density, must be considered as defects may act as trapping sites for hydrogen and, thus, reduce the concentration of diffusible hydrogen [19]. In comparison to conventionally produced steel, PBF material exerts a higher dislocation density in the as-built state [20]. Contrary, thermal postprocessing of PBF material by solution annealing reduces dislocation density by recovery and, at higher temperatures, recrystallization [14].

The absorption of hydrogen within the crystal lattice of a steel is accompanied by changes in physical properties not only with respect to mechanical properties like toughness and ductility: hydrogen incorporation also results in subtle changes in the electronic and magnetic properties. Each hydrogen atom introduces an additional electron and spin into the material and, thus, also changes the local electronic and magnetic structure being of particular importance in Fe-base materials [21, 22]. An argument based on numbers underestimates the influence of hydrogen. A few 10 ppm more or less electrons do not make much difference in the total number of electrons in a metal. However, it has recently been shown that steel samples loaded with different amounts of hydrogen differed significantly in their Seebeck coefficients [23]. Here, the Seebeck coefficient represents a sensitive probe for changes in the steepness of the electronic density of states near the Fermi edge [24]. Therefore, our hypothesis is that these measurement results can be explained by the property of hydrogen to attach directly to electrically and – in steel – usually also magnetically active crystal defects. This can very effectively modify the steepness of the electronic density of states near the Fermi edge.

Based on this hypothesis, in this work we investigate the effects of hydrogen incorporation on the subtle changes in the electronic and magnetic properties, using differently prepared microstructure states. Three X2CrNi18-9 stainless steel samples of identical global chemical composition but different manufacturing routes are probed for this purpose. This offers the possibility to investigate the impact of hydrogen incorporation in dependence of homogeneity, grain size, and dislocation density. The electronic properties of the samples are examined using Seebeck coefficients – a transport coefficient that reacts very sensitively to small changes in the electronic density of states at the Fermi edge. The magneto-structural properties are probed by a combination of diffuse neutron scattering and magnetic small angle neutron scattering (SANS).

## 2. Results

**2.1 Microstructural characterization**

For our study, we selected a metastable austenitic steel within the compositional range of DIN EN X2CrNi18-9 (AISI 304L), cf. Table 1. To investigate the role of chemical homogeneity on the effects of hydrogen absorption, we created a matrix of three prepared material states. For each state, we subjected one sample to a hydrogen atmosphere and one reference sample to a pure argon atmosphere. The three prepared material states were

(1) conventional processing by casting and forging followed by a final solution annealing step (**CON-SA**),

(2) laser powder bed fusion also in combination with solution annealing (**PBF-SA**),
(3) laser powder bed fusion in the as-built condition without further thermal treatment (**PBF-AB**).

Table 1: Mean alloy compositions of DIN EN X2CrNi18-9 in three different processing conditions measured by optical emission spectrometry; values given in mass-%.

| Condition | C | Si | P+S | Cr | Ni | Cu | N | Fe |
|---|---|---|---|---|---|---|---|---|
| CON-SA | 0.06 ±0.001 | 0.5 ±0.03 | <0.03 | 17.1 ±0.2 | 10.4 ±0.1 | 0.8 ±0.02 | 0.03 ±0.001 | bal. |
| PBF-SA | 0.03 ±0.001 | 0.4 ±0.01 | <0.03 | 17.3 ±0.1 | 10.1 ±0.1 | 0.8 ±0.01 | 0.06 ±0.002 | bal. |
| PBF-AB | " | " | " | " | " | " | " | " |

Table 2: Hydrogen content after charging, measured by carrier gas hot extraction; values are given in weight-ppm

| Condition | H |
|---|---|
| CON-SA | 12.78 ±0.68 |
| PBF-SA | 10.80 ±0.01 |
| PBF-AB | 11.03 ±0.44 |

Exemplary BSE (backscattered electron) images of these samples were published and discussed in [16] and are shown in Figure 1a-c. Becker et al. [16] report that Figure 1c, depicting the PBF-AB microstructure parallel to the build direction, reveals columnar grains oriented along the build direction. Compared with the CON-SA (Fig. 1a) state, the PBF-AB condition is distinguished not only by its columnar grain morphology but also by a markedly smaller mean grain size of 9.5 µm. In contrast, CON-SA exhibits substantially larger average grain size of 27.7 µm and displays the typical microstructure of conventionally produced austenitic steels, i.e., predominantly equiaxed grains with occasional annealing twin boundaries. Moreover, inspection of the PBF-SA micrograph (Fig. 1b) with an average grain size of 8.2 µm indicates that a subsequent 1 h heat treatment at 1050 °C does not significantly alter the grain structure of the PBF-AB condition [16].

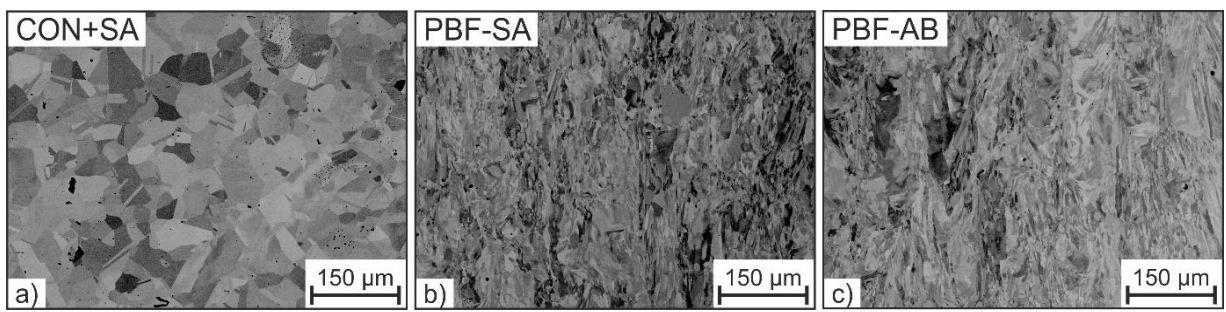

*Figure 1: Comparison of X2CrNi18-9 steel microstructures across three processing conditions (CON-SA, PBF-SA, PBF-AB) by BSE imaging; Images a)–c) adapted from Becker et al. [16], Metallurgical and Materials Transaction A (2026), DOI: 10.1007/s11661-025-08102-x, CC BY 4.0; changes: rearranged.*

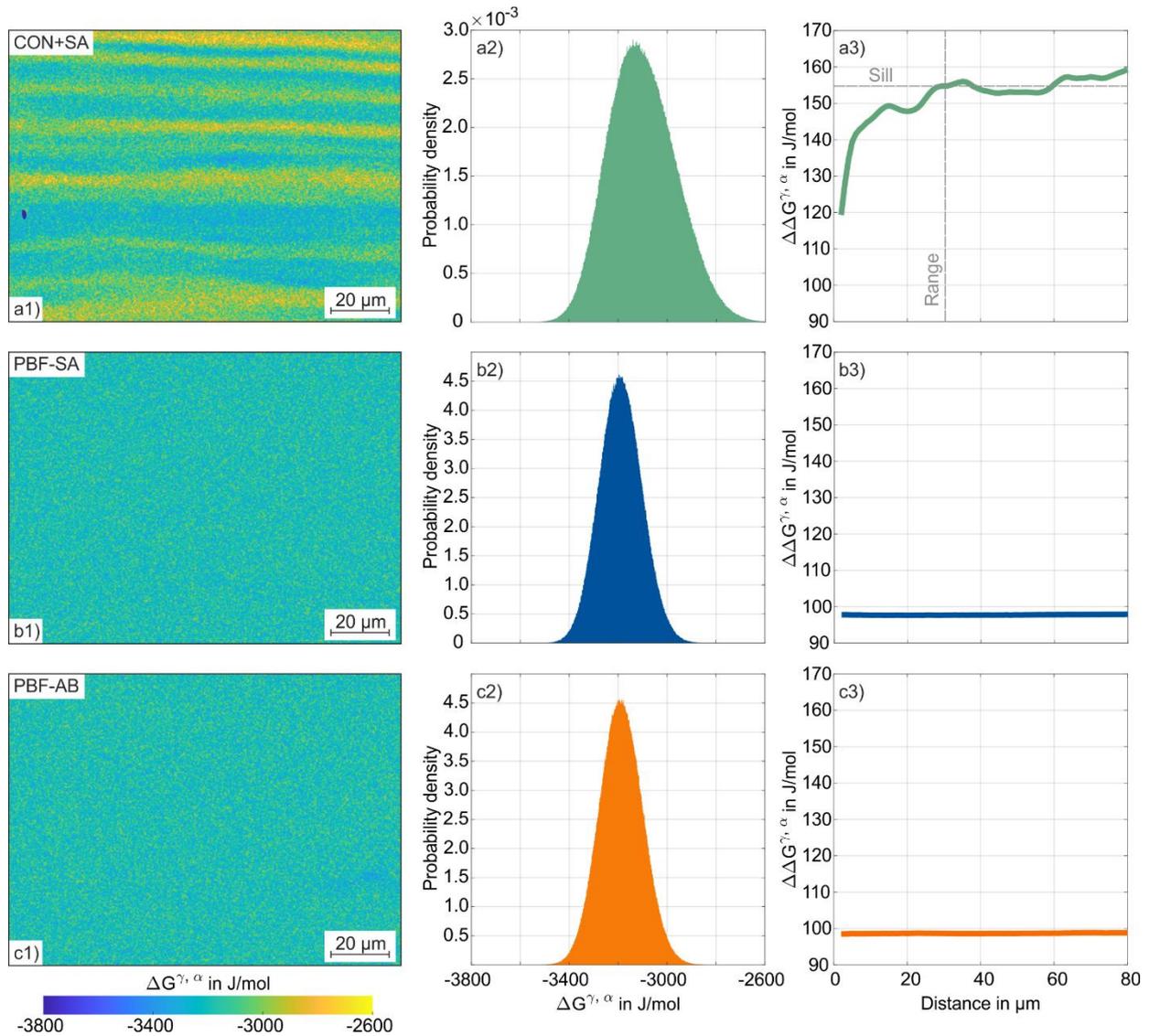

*Figure 2: Comparison of the mesoscale chemical homogeneity of the studied specimens a1-3) CON-SA, b1-3) PBF-SA c1-3) PBF-AB. a1), b1), c1) Phase stability maps (ΔG maps) computed with Thermo-Calc (TCFe10 database) and based on elemental maps measured by EDS. a2), b2), c2) histograms of the values in the ΔG maps. a3), b3), c3) First-order variograms of the ΔG maps.*

In addition to the grain structure, Becker et al. [16] investigated the elemental distribution in the different conditions using EDS (Energy Dispersive Spectrometry) mapping. These datasets were converted into phase stability maps (ΔG maps) using the Thermo-Calc/Python SDK TC-Python coupled with the thermodynamic database TCFe10. The Gibbs free-energy difference, ΔG, between austenite (γ phase) and ferrite (α phase) serves as an indicator of the thermodynamic stability of austenite, i.e., its propensity to transform into bcc martensite either thermally or under applied stress or strain. Figure 2 presents the ΔG maps (a1-c1), the corresponding probability density normalized histograms (a2-c2), and the first-order variograms (a3-c3) derived from the elemental distributions.

The CON-SA condition exhibits pronounced local variations in ΔG. This heterogeneity is quantitatively confirmed by the histogram in Fig. 2 a2), which displays a significantly broader distribution (standard deviation: 141 J/mol) compared to the narrower distributions of the additively manufactured PBF-AB (std: 87 J/mol) and PBF-SA (std: 88 J/mol) conditions.

It is worth noting that the mean ΔG values also differ slightly between the processing routes: -3100 J/mol for CON-SA versus approximately -3190 J/mol for the PBF conditions. Considering that the starting material for all specimens was the same [16], these shifts likely stem from slight variations in chemical composition induced by the different processing methods, as detailed in Table 1. While the conventionally produced material (CON-SA) has a higher carbon content (0.06 wt.%) compared to the PBF material (0.03 wt.%), the nitrogen content shows the inverse trend, increasing from 0.03 wt.% in CON-SA to 0.06 wt.% in the PBF conditions. Despite these compositional shifts, the relative difference in mean stability is small (< 3%), allowing the materials to be regarded as thermodynamically comparable for this analysis.

To quantify the spatial structure of the chemical gradients, first-order variograms were calculated (Fig. 2 a3-c3). These display the mean absolute difference between value pairs in the maps as a function of the distance between them [25]. For the CON-SA condition (Fig. 2 a3), the variogram reveals the spatial dependency of the segregation. The curve rises to a stable plateau (the sill) of approximately 155 J/mol. This sill represents the mean absolute difference between any two statistically independent points in the field (see experimental section). The range, i.e., the distance at which this sill is reached, is approximately 30 µm. This value corresponds to the characteristic width of the band-shaped variations visible in the ΔG map (Fig. 2 a1).

In contrast, no discernible local differences are observed for the powder-metallurgically produced conditions PBF-SA and PBF-AB. The histograms (Fig. 2 b2, c2) are narrower, and the corresponding first-order variograms (Fig. 2 b3, c3) appear as flat lines at a constant level of 98 J/mol. This lack of spatial correlation indicates that the mean absolute difference between points is independent of distance, confirming chemical homogeneity at the resolved scale. Quantitatively, the mesoscopic segregation in the cast material (represented by the sill of 155 J/mol) results in a mean absolute difference that is roughly 58% higher than the baseline variation observed in the PBF materials. It should be noted that the nugget effect (the fact that the curves do not start at the origin) observed in all specimens is a result of the median filtering stage applied during data processing (cf. experimental procedures).

The mentioned banding in CON-SA is attributed to Ni-rich and Ni-lean bands. A corresponding band-like segregation pattern of Fe and Cr was also observed, showing an inverse behavior compared with Ni: regions with lower Ni concentrations exhibited higher measured Fe and Cr contents [16]. This opposing segregation is explained by the casting solidification path, which proceeds via primary ferrite. Since ferrite has an inherently higher solubility for Cr and a lower solubility for Ni, Ni partitions into the remaining liquid, promoting austenitic solidification of the residual melt.

The reason why no local variations in ΔG or elemental distributions are observed for the PBF-SA and PBF-AB conditions is the small melt volumes present during powder atomization and powder bed fusion-laser beam melting (PBF-LB/M) fabrication. These small melt pools lead to segregation structures with much smaller characteristic dimensions. While PBF-LB/M generates a cellular substructure with increased Ni concentration at the cell boundaries on the sub-micrometer scale [26], these features are not resolved at the spatial scale considered here. Subsequent solution annealing (PBF-SA) homogenizes these chemical gradients due to the short diffusion distances involved [27].

The final microstructural parameter addressed here is the dislocation density. To quantify it, the X-ray diffractograms previously described by Becker et al. [16] and shown in Figure 3 were analyzed with respect to the dislocation density (see Experimental for further details). Table 3 summarizes the dislocation densities for the different states.

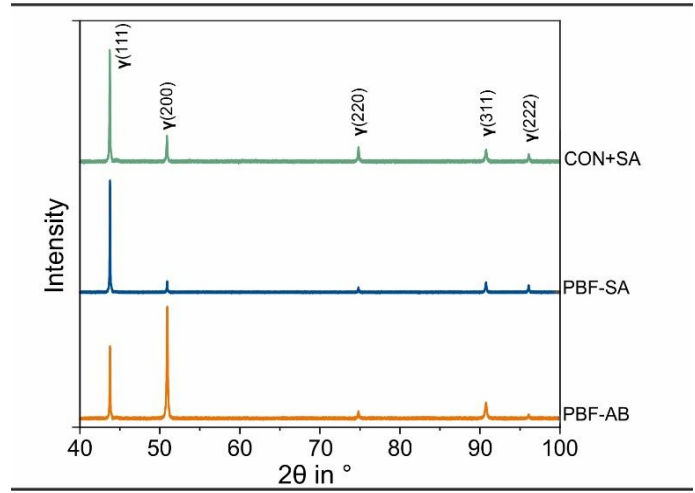

*Figure 3: X-ray diffraction patterns of the investigated states. Image adapted from Becker et al. [16], Metallurgical and Materials Transaction A (2026), DOI: 10.1007/s11661-025-08102-x, CC BY 4.0; changes: cropped, rearranged, and color-coded.*

***Table 3***: Dislocation densities obtained from the X-ray diffraction patterns shown in Figure 3.

| Condition | $\rho$ in $1/m^2$ |
|---|---|
| CON-SA | $3.04 *10^{13}$ |
| PBF-SA | $2.97 *10^{13}$ |
| PBF-AB | $3.88 *10^{13}$ |

It becomes evident that the dislocation density is highest in the PBF-AB condition. This can be attributed to the process-inherent thermal conditions, characterized by the formation of small melt pools that solidify and cool rapidly. This, in turn, promotes high thermal stress, which is relieved by microplastic deformation. These mechanisms lead to the development of an increased dislocation density [28]. After subsequent solution annealing, the dislocation density decreases markedly in the PBF-SA condition. This reduction is attributed to recovery processes, as has been shown in several studies [29, 30]. A similar range is observed for the CON-SA condition, suggesting that post-solution annealing results in comparable dislocation densities, irrespective of whether the material was produced by conventional or additive manufacturing.

### 2.2 Electronic and magneto-structural properties

To understand how the processing routes and consequent microstructures influence electronic and magneto-structural properties with and without hydrogen incorporation, we measured the Seebeck coefficient, diffuse neutron scattering and small-angle neutron scattering (SANS). Taking into account the different sample geometries, the samples were loaded in an autoclave at 300 °C and a pressure of 10 bar in 100% hydrogen gas for as long as necessary to ensure complete and homogeneous incorporation of hydrogen in each case. Reference samples were subjected to the same process, but with argon gas. The temperatures for this incorporation were selected so that microstructural processes would not be dominant. The thermoelectric voltage was measured in a direct measurement protocol from 300 K to 345 K. The heating and cooling curves were measured for each material state in order to monitor differences, for example due to hydrogen effusion. Results are shown in Fig. 4.

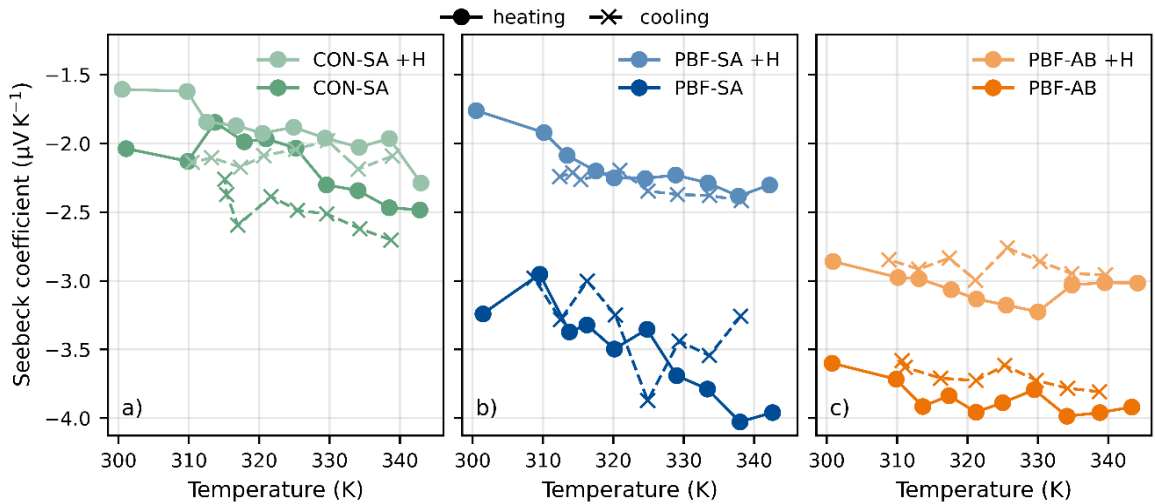

*Figure 4: Temperature-dependent Seebeck coefficient S in the range 300–345 K CON-SA, PBF-SA, and PBF-AB. Measurements are shown for the hydrogen-free and hydrogen-incorporated conditions. Heating and cooling cycles are indicated by solid and dashed lines, respectively.*

All samples have negative Seebeck coefficients. This means that electrons are the main charge carriers, as expected for this steel composition [31]. In general, the lower the absolute value of the Seebeck coefficient, the more metallic the sample is, i.e., the higher its concentration of conduction electrons [32]. Differences between the heating and cooling curves in the temperature range between 300 K and 345 K are minor.

The differences between the three samples with identical global chemical composition but different processing are evident. The conventionally processed sample CON-SA and the samples processed by PBF, PBF-SA and PBF-AB, differ in the absolute value of the Seebeck coefficients, even in the reference condition where no hydrogen was incorporated. We conclude that the PBF samples have a distinct electrochemical potential compared to the conventionally processed ones. While the microstructures of these samples differ significantly (see Fig. 1), it is likely that the chemical inhomogeneities at the mesoscale (see Fig. 2) cause the different Seebeck coefficients, since changes in chemical composition directly cause changes in electrochemical potential. Notably, the trends in the absolute Seebeck coefficients of the samples do not correlate with differences in dislocation density.

In all cases, the hydrogen-treated samples show measurable differences compared to the reference samples: the absolute values of the Seebeck coefficient are smaller than the values of the argon-treated samples. All samples have similar values of incorporated hydrogen in the order of 10 ppm (see Table 2). Given that the samples approximately contain electrons in the order of $10^{22}$ to $10^{23}$ cm$^{-3}$, this means changes in the order of a maximum of $10^{19}$ cm$^{-3}$ due to hydrogen incorporation. Therefore, measurable changes in a transport coefficient in the percentage range cannot be explained simply by the additional number of charge carriers. We conclude that there should be an electronic mechanism at work, which warrants further study and potentially also could be used sensing principle based on Seebeck measurements as proposed in Ref. [23].

Interestingly, the values of the two samples, which were processed differently but both underwent identical thermal post-treatment (solution annealing), CON-SA and PBF-SA, are similar when these samples have hydrogen incorporated. This is particularly noteworthy, as these two samples differ significantly in terms of their microstructure (Fig. 1) and their chemical inhomogeneity and austenite stability (Fig. 2). Nevertheless, these samples are comparable in terms of their electronically effective properties like their observable Seebeck coefficients after they have been saturated with hydrogen.

Earlier, we argued that the mesoscale inhomogeneities (Fig. 2) probably correlate with the different Seebeck coefficients of the samples, CON versus PBF. In this respect, we conclude now that hydrogen incorporation seems to compensate for these differences, at least in the case of the solution annealed state, and that the sample states seem to converge here.

From the electronic properties, represented by the Seebeck coefficient, it can therefore be concluded that chemical inhomogeneities interplay with hydrogen incorporation in the samples. To determine the extent to which chemical homogeneity is truly a relevant parameter for hydrogen uptake in the material, we conducted further magnetic structural analyses on this series of samples.

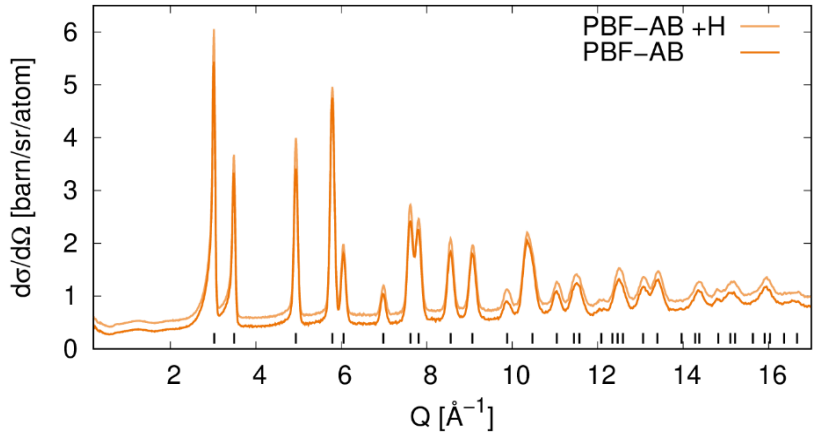

*Figure 5: Diffuse magnetic scattering of the PBF-AB processed sample with and without hydrogen incorporation. Theoretical reflection positions for the austenite phase (FCC) are marked.*

To investigate this question further, we first examined the PBF-AB sample with regard to its microscopic structural properties, both with and without hydrogen incorporation using diffuse neutron scattering at the NIMROD instrument (RAL-ISIS) [33]. The data presented in Figure 5 confirms the pure austenite (FCC) structure type for both samples. A significantly enhanced scattering background in the hydrogen loaded state indicates the nuclear-incoherent scattering induced by hydrogen in this sample. In addition, a small variation in relative reflection intensities is observed. While approximately 10 ppm hydrogen in a sample naturally enhances the scattering background due to hydrogen's large neutron scattering cross-section for neutrons, a variation in the relative reflection intensity is actually a highly relevant finding.

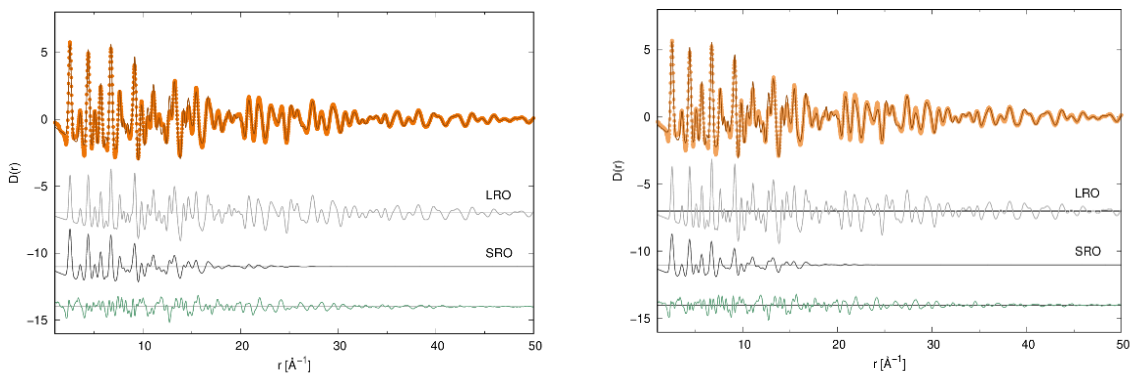

*Figure 6: Analysis of the differential pair correlation function D(r) of PBF-AB without (left) and with (right) H treatment. Neutron PDF data (orange) is shown with fit (dark orange), contributions by the*

*long-range (light grey) and short-range (dark grey) ordered phases, and the difference between data and fit (green).*

The differential pair correlation function D(r) obtained by Fourier transform of the data, shown in Figure 6, is in excellent agreement with the austenite phase. Considering grain sizes in the µm length scale expected for all samples, we refined the long-range order (LRO) as a bulk phase in the range of r > 30 Å. In the lower r range, we observe additional correlations that can only be described by a short-range ordered (SRO), nanoscale phase that is introduced to the fit as a second phase. This phase has the same structure type (fcc) as the bulk phase, but a correlation length of only 3.0(4) nanometres. Interestingly, when hydrogen is incorporated into the sample, the correlation length of the nanophase is slightly smaller with 2.7(4) nm. We attribute the found nanoscale phase to the nanoscale segregations typically found in the cell boundaries of PBF processed materials [26, 27].

To further understand the impact of hydrogen incorporation on the nanoscale structure and magnetism, all samples were characterized using Small-Angle Neutron Scattering (SANS). Magnetic SANS is a versatile technique to provide unique information on both structural and magnetic inhomogeneities on mesoscopic length scales (1-1000 nm) in bulk materials [34, 35]. It is therefore directly sensitive to nanoscale defects and disorder induced by grain boundaries, texture, and local defects as well as the arising spin misalignment. In the approach to saturation technique [36], the sample is magnetized to a nominally saturated state, where the spin misalignment in the sample is suppressed and mainly nuclear and collinear magnetic scattering contributions are observed. With decreasing magnetic field towards coercivity, the spin misalignment scattering contribution increases and becomes the dominant scattering contribution. Magnetic SANS is particularly sensitive to the spin misalignment for materials with weak nuclear scattering, such as chemically homogeneous bulk ferromagnets.

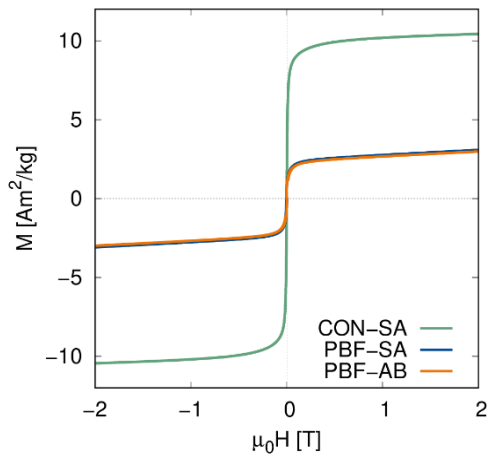

*Figure 7: Mass magnetization of the studied samples without hydrogen treatment.*

The samples under study consist of austenitic steels and are considered macroscopically chemically homogeneous. Austenite is typically a paramagnetic material at room temperature, however, a soft ferromagnetic signature is observed (Fig. 7), in agreement with previous reports [37]. The saturation magnetization in the PBF samples is significantly reduced as compared to the CON-SA sample, as can be expected for the higher defect density in the additively manufactured samples. We consider the domains in saturation to be very large, beyond the length scales accessible to our SANS experiment and hence leading to negligible magnetic contrast in the collinear magnetization. In a nominally saturated state, we therefore expect only small contributions of a weak nuclear contrast (i.e., structural defects and distortions in the nanoscale) and a small contribution of remaining spin misalignment. In the absence of a magnetic field, corresponding to a fully disordered state, a significant amount of spin disorder may arise from grain boundaries and defects, leading to a significantly enhanced scattering response.

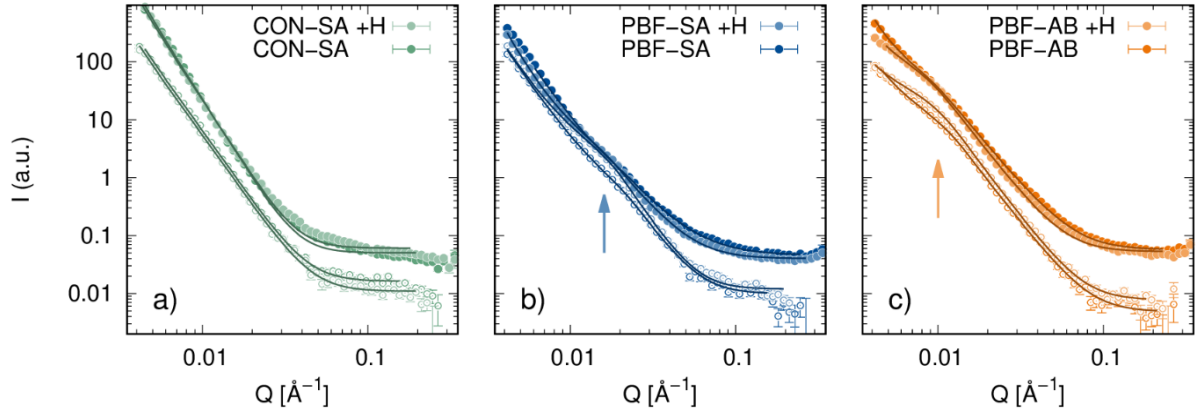

*Figure 8: Magnetic SANS collected in applied field of 1 T (empty symbols) and 0 T (full symbols) by austenite steels with (lighter color) and without (darker color) hydrogen treatment. a) Sample (1) prepared by conventional processing, b) sample (2) prepared by laser powder bed fusion (PBF) processing with solution annealing, and c) sample (3) prepared by PBF. Signatures of nanoscale inhomogeneities are marked (arrows).*

The magnetic SANS by the three samples under study in hydrogen-treated and Ar-treated conditions is presented in Figure 8. Data was acquired at the ZOOM instrument (RAL-ISIS) both in an applied magnetic field of 1 T and in a near-remanence state (0 T).

The SANS by sample CON-SA (Fig. 7a) follows a power law, which can be understood as the Porod scattering of large grains with length scales in the higher µm range that is not accessible by our experiment. The larger overall scattering intensity observed in the absence of a magnetic field is attributed to the spin disorder induced by grain boundaries, increasing the power law exponent from -3.9(1) in 1 T to -4.3(3) in 0 T. In this sample, the effect of hydrogen incorporation on the scattering response is negligible.

For the samples PBF-SA and PBF-AB (2) and (3), SANS reveals an additional signature of nanoscale inhomogeneities that is observed in the Q ranges of 0.01 – 0.02 Å$^{-1}$ and modelled using a Guinier law contribution [38]. A Guinier radius of $R_G$ = 17.9(2) nm is derived for the inhomogeneities in the PBF-AB as-built condition (sample 3). After solution annealing (PBA-SA), these inhomogeneities are significantly smaller, with $R_G$ = 9.4(2) nm, and much lower in number density. We attribute these inhomogeneities to the sub-micron segregations in the cell boundaries of PBF processed materials [26, 27], in line with the short-range order phase found on the atomistic scale, by analysis of the differential pair correlation function (Fig. 6). Upon annealing, their number and extent is significantly reduced.

Upon hydrogen incorporation, the observed inhomogeneities grow to $R_G$ = 20.3(1) nm in the PBF-AB state and $R_G$ = 12.2(1) nm in the annealed state (PBF-SA). We attribute this observation to a clear structural effect of hydrogen incorporation, such as an enrichment of hydrogen near the grain boundaries. In the fully disordered state without applied magnetic field, the signature of these inhomogeneities further grows for the as-built material (Fig. 7b, sample PBF-AB) from $R_G$ = 17.9(2) nm to 20.5(1) nm (without hydrogen incorporation) and from $R_G$ = 20.3(1) nm to 22.5(1) nm (with hydrogen incorporation). This behavior is attributed to the significant scattering contribution of spin disorder arising in the nanoscale grain boundaries, i.e., the structural inhomogeneities induce spin misalignment that reaches further into the material. For the annealed sample (sample PBF-SA), the size of the inhomogeneities is nearly independent on the applied field, which may indicate a lower degree of disorder in these already smaller and less prominent defects.

It is clearly visible in Figure 7, how the SANS by the hydrogen incorporated PBF samples (with and without annealing, PBF-AB and PBF-SA) in 1T and 0T approaches each other as compared to the non-loaded states. This indicates a lower amount of spin disorder after hydrogen incorporation, an effect that is even stronger in the solution-annealed sample (PBF-SA).

We therefore conclude from our magnetic SANS analysis that the PBF processed materials exhibit a significant amount of nanoscale inhomogeneities that are less prominent after solution annealing. These inhomogeneities are more pronounced after hydrogen incorporation, indicating that the hydrogen preferably collects near these nanoscale segregation structure in the cell boundaries of the material. The spin misalignment scattering contribution further indicates a lower degree of spin misalignment in the hydrogen incorporated samples, suggesting that hydrogen incorporation is more effective in reducing the degree of spin disorder in the annealed state. This is in line with the much stronger effect of hydrogen incorporation on the electronic properties of this material (Seebeck data analysis) as compared to the as-built PBF condition.

**Conclusion**

This study shows the effects of hydrogen incorporation on the electronic and magneto-structural properties of X2CrNi18-9 stainless steel in three different microstructural states: conventionally processed and solution annealed (CON-SA), additively manufactured by laser powder bed fusion (PBF) in the as-built state (PBF-AB) and in the subsequent solution annealed state (PBF-SA). The results show that hydrogen incorporation causes measurable changes in the electronic structure of all samples, which is reflected in a reduction in the absolute values of the Seebeck coefficients. This is particularly remarkable given that hydrogen was incorporated in the sample at a level of 10–12 ppm. This would change the electron density in the fifth decimal position. Clearly, there must be a very powerful mechanism at work for hydrogen to significantly alter the steepness of the electronic density of states near the Fermi energy. To understand this, we conducted further magneto-structural investigations. The magneto-structural properties are also altered by the incorporation of hydrogen. In particular, evidence of the formation of short range order nanoscale inhomogeneities in the samples was found. Same as the bulk phase, these inhomogeneities have a fcc structure, but with correlation lengths of only 3 nm. We find that hydrogen preferentially accumulates in the vicinity of these nanoscale inhomogeneities. This was also evident in a reduction in spin disorder due to the incorporation of hydrogen. Future research should therefore focus on further investigation of these nanoscale cell boundaries, as they appear to be crucial for the incorporation of hydrogen in the austenitic steel. With this knowledge, the composition and length scales of the cell boundaries might be tuned by PBF processing parameters, to rationally optimize the hydrogen uptake efficiency. Furthermore, extending this study to other steel compositions and processing routes could further advance the understanding of hydrogen uptake in stainless austenitic steels and as a next step guide the development of materials with improved hydrogen-resistance. Furthermore, the observed changes in the measured Seebeck coefficients could potentially serve as sensitive probes for non-destructive investigation of hydrogen incorporation within these materials.

**Experimental**

**Material:**

The material was produced using three different manufacturing routes, as detailed in Ref. [16]. The wrought material was manufactured by melting a 3 kg ingot in a vacuum induction furnace under an argon atmosphere at 500 mbar. The ingot was then hot-forged at 1100 °C with a round-die hammer in six forging passes, reducing the diameter from 42 mm to 12 mm. This procedure resulted in a forming strain of 42% per pass and a total cumulative strain of 251%. The chemical composition was determined by spark optical emission spectrometry (OES).

For the powder-metallurgical processing route, powder was produced using a Blue Power vacuum induction melting inert gas atomizer. The feedstock was melted under argon at 1.02 bar and 1710 °C. Atomization was performed with nitrogen at 30 bar, without preheating. Following atomization, the powder was sieved and separated into particle-size ranges of 20–63 µm for PBF-LB/M.

PBF-LB/M fabrication was conducted on a TruPrint3000 machine (TRUMPF GmbH) using the following parameters: 160 W laser power, 900 mm/s scan speed, 30 µm layer thickness, 80 µm hatch spacing, nitrogen ($N_2$) as the process gas, and a stripe scan strategy. These parameters were selected based on a prior parameter study targeting the maximum achievable density in the manufactured specimens.

Solution annealing was performed in an industrial vacuum furnace (U 54/1 F, Schmetz) at 1050 °C for 60 min, followed by argon gas quenching at 200 kPa resulting in the two conditions CON-SA and PBF-SA. The third condition considered in this work, PBF-AB, did not undergo any further heat treatment, as already mentioned in the Introduction.

**Microstructural characterization:**

The chemical and resulting phase stability homogeneity were characterized through analytical electron microscopy. Based on elemental maps obtained by energy dispersive spectrometry (EDS), thermodynamic stability maps of the present austenite phase were computed. At each individual position in the obtained field, the local chemical composition was employed to conduct two single equilibrium Calphad calculations: One to determine the Gibbs free energy of the austenite (fcc, A1), and one to compute the Gibbs free energy of the ferrite phase (bcc, A2). These were then subtracted at each position, to ultimately create a real-valued 2D map of the chemical stability of the austenite phase. We have then employed this so-called ΔG descriptor as an indicator of the stability of the austenite phase against a thermally or mechanically induced transformation into martensite [39]. It is worth noting that the here-employed ΔG corresponds to the chemical driving force for the transformation; while several methods exist to account for the total (or net) driving force for the transformation of austenite into martensite in Fe-based materials, we decided against them in the interest of conciseness.

The Calphad calculations were conducted employing the commercially available software Thermo-Calc 2025a through its SDK TC-Python. Each map consisted of 1345 x 1793 positions, i.e., 2,411,585 individual locations, totalling almost five million single equilibrium calculations. To speed the process, batch calculations and parallelization were employed. A median filtering stage was subsequently applied to the data to reduce noise.

To quantitatively analyze the obtained thermodynamic data, we employed histograms and first-order variograms [25]. As opposed to histograms, variograms, a tool originally conceived in geostatistics to analyze soils, convey spatial correlation information. First-order variograms evaluate the mean absolute difference (MD) between pairs in a field as a function of the distance and orientation of the vector between them. In this work, we present them as radially averaged variograms, which can be plotted in cartesian axes. In this representation, the horizontal axis corresponds to the spatial distance between pairs of points (all directions are averaged), while the vertical axis displays the mean absolute difference in ΔG values calculated for those pairs. In such a representation three main features of variograms become specially interesting: the sill, the range, and the nugget effect. If there are spatial correlations between the values in the analyzed field, these appear as deviations from a straight line.

The sill, if it exists, is the constant MD obtained when no spatial correlations can be observed. The range is the distance at which the sill is reached. The nugget effect, finally, is the jump between the first point in the variogram curve and the origin; in this study, this effect is the result of the applied median filtering stage.

To obtain qualitative indications of the dislocation density in the different states, X-ray diffractograms already shown in [16] were analyzed. Rietveld refinement was performed using the MAUD (Materials Analysis Using Diffraction) software, based on the Rietveld method [40, 41]. This analysis was used to determine the lattice parameter $a$, the mean crystallite size $D$, and the microstrain $\varepsilon$. These parameters were then used to calculate the dislocation density $\rho$ according to the Williamson–Smallman relation [42]:

$$\boldsymbol{\rho = \frac{2\sqrt{3}\epsilon}{bD}}.$$

**Hydrogen charging:**

All samples were hydrogen-charged in a high-pressure autoclave (Buchi Novoclave, 773 K / 500 bar rating). Prior to charging, the autoclave chamber was purged three times with argon to remove residual air and moisture. For hydrogen loading, the autoclave was filled with 100 % $H_2$ to an initial pressure of 10 bar at room temperature. The temperature was subsequently increased to 573 K with an average heating rate of approximately 2–3 K min$^{-1}$, corresponding to a heating time of roughly 1.5–2 h. This temperature was selected to ensure sufficiently fast hydrogen uptake while avoiding microstructural changes unrelated to hydrogen.

The below reported charging times were counted from the moment the target temperature was reached. After this point, the temperature required additional time to fully stabilize due to the thermal inertia of the system; this stabilization phase is included in the reported charging times and was not treated separately.

For samples intended for Seebeck coefficient measurements, specimens were charged at 573 K for 72 h. In contrast, samples prepared for diffuse neutron scattering and magnetic SANS measurements were charged for 50 h due to their smaller geometry, ensuring comparable saturation levels across these techniques. Specimens for hydrogen quantification via carrier gas hot extraction were charged for 80 h. All charging times guaranteed complete and homogeneous hydrogenation based on diffusion coefficients from literature [43], with the extended duration for extraction samples providing additional margin for uniform saturation.

During heating, the gas pressure increased according to the autoclave's temperature–pressure behavior (approximately ideal-gas-like expansion), while no active pressure regulation was applied. After the respective loading period, the autoclave was cooled to room temperature at its intrinsic cooling rate and subsequently depressurized before sample removal.

To distinguish hydrogen-induced changes from those caused solely by thermal exposure, we prepared reference samples under identical conditions but with argon atmosphere instead of hydrogen. For this purpose, the autoclave was filled with 7 bar Ar, and the same heating ramp, maximum temperature and dwell time were applied. These samples serve as a reference and allow identification of purely hydrogen-related effects.

Immediately after removal from the autoclave, the hydrogen-charged samples intended for Seebeck measurements were rapidly quenched and stored in liquid nitrogen in order to minimize hydrogen effusion. The samples were kept under cryogenic conditions until shortly before the Seebeck measurements. Prior to mounting in the measurement setup, the samples were allowed to thaw in ethanol to ensure a controlled transition from cryogenic to ambient temperature and to prevent ice formation on the sample surface. This procedure ensured that the measured thermoelectric response reflects the hydrogen-charged state as accurately as possible.

**Seebeck measurement:**

The Seebeck coefficient was measured using a ULVAC Riko ZEM-3 system in the steady-state direct-method configuration. The samples had a cross-section of approximately 2.0 mm × 2.0 mm and a length of 10 mm and they were mounted with a thermocouple spacing of 3.0 mm. Graphite paper was used between the samples and the electrodes to ensure reproducible electrical and thermal contact while preventing chemical interaction with the electrode blocks.

Before the temperature sequence was started, a room-temperature measurement was performed. The sample temperature was then varied from 303 K to 345 K in increments of 5 K, followed by a symmetric cooling sequence back to 303 K. At each temperature point, three different temperature gradients ($\Delta T$) were applied to determine the Seebeck coefficient accurately. The actual temperature differences were measured by the internal thermocouples of the ZEM-3 system.

For each temperature point, the Seebeck coefficient $S$ was determined using the standard ZEM-3 direct-method evaluation [44]. In this approach, the thermoelectric voltage V generated across the sample is measured as a function of the temperature difference $\Delta T$ applied along the sample length. The Seebeck coefficient was obtained from the slope of the linear relationship between $V$ and $\Delta T$, according to [45]:

$$S = -\frac{dV}{dT} \qquad .$$

A complete Seebeck measurement sequence, including heating and cooling between 303 K and 345 K, required on the order of 7 h per sample. This measurement duration is relevant with respect to possible hydrogen effusion during the experiment.

**Neutron scattering:**

Hydrogen-incorporated specimens for neutron scattering experiments underwent a 10-day period of transit and storage at the beamline facility under ambient conditions without active cooling prior to measurements. This protocol is justified by the characteristically low hydrogen desorption kinetics of austenitic steels at room temperature [43], ensuring minimal hydrogen loss during the exposure window.

Neutron total scattering (neutron PDF) was measured on the Near and Intermediate Range Order Diffractometer (NIMROD) at the ISIS pulsed source [33]. NIMROD views a wide range of wavelengths, from epithermal to cold neutrons from a water pre-moderator and liquid hydrogen moderator (0.05-14 Å). This is coupled with a wide forward scattering detector array (0.5-40 degrees) affording a uniquely wide Q-range (0.02-50 Å$^{-1}$). The samples were mounted in Vanadium foil packets perpendicular to the beam and measured at room temperature in the instrument vacuum tank for a minimum of 5 hours. A beam size of 30x30 mm was used, larger than the 7.8 mm diameter samples, with the data normalized on a per atom basis using the mass of composition of the sample. Data were reduced using the Gudrun software [46], which subtracts scattering from the empty V foil container and the empty instrument, normalizes to a VNb null coherent scattering standard, performs multiple scattering and inelasticity corrections and Fourier Transforms the data to real-space pair distribution functions.

The obtained differential pair correlation function data was evaluated using the PDFgui program [47].

Unpolarized magnetic SANS data was measured at the ZOOM instrument (ISIS, UK) with and without a horizontally applied magnetic field (perpendicular to the neutron beam) of 1 T. A detector distance of 4 m was used and a neutron wavelength band of 4.0-16.5 Å. The data reduction was performed using the Mantid software [48]. A background scattering contribution of Al foil was subtracted. The radially integrated data was modelled according to a Guinier-Porod model [38] available in Sasview [49], in combination with a Porod law accounting for the presence of large, micron-sized grains.

**Declaration of Generative AI and AI-assisted technologies in the writing process**

During the preparation of this work, the authors used generative artificial intelligence (AI)–assisted tools to improve the readability and language of the manuscript. After using these tools, the authors carefully reviewed and edited the content as needed and take full responsibility for the content of the published article.

**Acknowledgement**

We gratefully acknowledge the Science and Technology Facilities Council (STFC) for access to neutron beam time at ISIS using the NIMROD (2590198) and ZOOM (2590187) facilities. This work benefited from the use of the SasView application, originally developed under NSF award DMR-0520547. SasView contains code developed with funding from the European Union's Horizon 2020 research and innovation programme under the SINE2020 project, grant agreement No 654000.

Further, the authors gratefully acknowledge financial support of the German Research Foundation within the research project "Development of methods for the alloy design of austenitic steels with a homogeneous deformation behavior and a high resistance against hydrogen embrittlement" (grant no 56380481) and the CRC/TRR 270 (grant no 405553726).